\documentstyle[12pt,aaspp4]{article}  
\newcommand\lsim{\lower.5ex\hbox{$\; \buildrel < \over \sim \;$}}
\newcommand\gsim{\lower.5ex\hbox{$\; \buildrel > \over \sim \;$}}
\newcommand{\etal}{\mbox{ et~al.}}
\newcommand{\kms}{\mbox{ km~s$^{-1}$}}

\newcommand{\ciii}{CIII$\left.\right]$}

\begin{document}

\title{The detection of the $\lambda$ 2175 \AA\ feature, and further 
analysis of the BAL line profile structure in the gravitational lens
candidate UM425 
\footnote{Observations reported here were made with the Multiple Mirror 
Telescope Observatory, which is operated jointly by the University of Arizona
and the Smithsonian Institution}
}
\author{Andrew G. Michalitsianos}
\affil{NASA/GSFC Laboratory for Astronomy \& Solar Physics\\
Code 680, Greenbelt, MD 20771 }
\author{Emilio E. Falco}
\author{Jos\'e A. Mu\~noz}
\affil{Harvard-Smithsonian Center for Astrophysics\protect \\ 
       60 Garden Street \protect \\
       Cambridge MA 02138 }

\author{Paul Hintzen}
\affil{NASA/GSFC Laboratory for Astronomy \& Solar Physics\\
Code 681, Greenbelt, MD 20771 }

\author{Demosthenes Kazanas}
\affil{Laboratory for High Energy Astrophysics\\
Code 661, Greenbelt, MD 20771}

\authoremail{falco@cfa.harvard.edu, jmunoz@cfa.harvard.edu}
\vskip 1.0truecm

\begin{abstract}

We obtained MMT spectra of the gravitational lens candidate UM425 to compare
the redshifts and line profile structures of lens components A and B, which are
separated by $\sim$ 6\farcs5.  The  CIV $\lambda$ 1550 \AA\ emission in both A
and B exhibits Broad Absorption Line (BAL) structure, consistent with the
earlier detection of BAL structure in OVI $\lambda$ 1033 \AA\  and NV $\lambda$
1240 \AA\ that was found with the International Ultraviolet Explorer (IUE) in
component A.  Cross-correlation of the spectra  of A and B using emission lines
of CIV $\lambda$ 1550 \AA, HeII $\lambda$ 1640 \AA, NIII $\lambda$ 1750 \AA,
\ciii\ $\lambda$ 1909 \AA\ and MgII $\lambda$ 2800 \AA\  reveals a difference
in the  redshifts of A \& B. However, the detailed BAL line profile structure
found in spectra of A \&  B are strikingly similar to one another, which
suggests the system is lensed. The spectra of A \& B also indicate significant
dust extinction, which we base on the presence of the $\lambda$ 2175 \AA\
absorption feature in the rest frame of the QSO (z$_{QSO}=1.47$). This feature
is commonly seen in galactic sources, but is not generally observed in QSO
spectra. Our spectra show the presence of the $\lambda$ 2175 \AA\ absorption
feature in spectra of both images associated with the gravitational lens UM425. 
Based upon the strong similarity of BAL line profile structure exhibited by
UM425A\&B, particularly the presence of the $\lambda$2175 dust absorption
feature in spectra of both images, we conclude that UM425 is a gravitational
lens.

\end{abstract}
\keywords{Gravitational~Lenses --- Dark~Matter ---
          Quasars:~individual~(UM425) }

\vfill
\eject

\section {Introduction}

The wide angular separation found in a modest population of
gravitationally lensed quasars suggests lensing by foreground
clusters. The gravitational lens candidate UM425 $= 1120+019$ consists
of two widely separated images ($\sim$ 6\farcs5) which were first
detected by Meylan \& Djorgovski (1988, 1989). A suspected foreground
cluster at z$\sim 0.6$ is a possible lens candidate, although evidence
for its existence was not found in the Hubble Space Telescope (HST)
Snap Shot Survey (Maoz \etal\ 1993). The large difference in luminosity
between components of 5.6 mag (m$_{V A} = 16.2$, m$_{V B} = 21.8$) 
corresponds to a large relative magnification factor of $\sim$100.

Spectra of UM425A that Michalitsianos \etal\ (1995, 1996) obtained
with the International Ultraviolet Explorer (IUE) indicate pronounced
Broad Absorption Line (BAL) structure in OVI $\lambda$ 1033 \AA\ and
NV $\lambda$ 1240 \AA, consistent with peak outflow velocities that
range up to $\sim-12000$ \kms.  As such, UM425 is a member of the
rare population of BAL QSOs with z$\gsim$1.5 that are gravitationally
lensed. A campaign to monitor the BAL high ionization line profile
structure over timescales $\sim$10 months with IUE indicated
significant changes in the OVI BAL absorption trough occur primarily
at velocities $\geq -4000$ \kms, while absorption features at
velocities $\leq -4000$ \kms\ were characterized by more modest
fluctuations.

Meylan \& Djorgovski (1989) estimated z$_{QSO} = 1.465$, based upon
the \ciii\ $\lambda$ 1909 \AA\ line. The IUE spectra yielded a slightly larger
z$_{QSO} = 1.471 \pm 0.003$, as estimated from the Ly$\alpha$ 
$\lambda 1216$ \AA\ line.  
The IUE value is indistinguishable from our optical estimate; see below.

\section{Data Analysis}

We obtained spectra of UM425 A\&B with the MMT and the Blue Channel
spectrograph. Table \ref{tab1} shows a log of the observations.  The useful
range of observed wavelengths was $3350-8000$\AA, with a resolution
of 1.96 \AA\ ${\rm pixel}^{-1}$ and an effective resolution (FWHM) of
7 \AA.  We calibrated the flux in the object spectra with spectra of
the spectrophotometric standards GB191B2B and PG1545+035 that we
acquired on the same night (see Table \ref{tab1}).
The calibrated MMT spectra for components A and B are overplotted in
Figure 1. These two spectra are a combination of 2 exposures,
2700 s each, for a total integration time of 1.5 hr.

We found a significant redshift difference between images A and B.
Using IRAF task emsao, we find the following redshifts for the two
components (based only on the high-SNR emission lines listed within
parentheses): $z_A=1.471 \pm 0.002$ (SiIV, CIV, HeII, \ciii\ 
and MgII), and $z_B=1.477 \pm 0.002$ (SiIV, CIV, HeII, NIII,
\ciii\ and MgII). To confirm the disagreement in these redshifts, we
cross-correlated the spectra with IRAF task xcsao, with component A in
the role of template; we obtained a well-defined peak in the
correlation for velocity $v=627.6$ \kms\ for component B with respect
to A.  To summarize, from the emission line measurements, $z_B-z_A =
0.006 \pm 0.003$; from the cross-correlation, $z_B-z_A = 0.005 \pm
0.001$.

We found two lines in component B, HeII  $\lambda$ 1640 \AA,
 and NIII 
$\lambda$ 1750 \AA\ with an amplitude that was very
different from that of component A, especially for NIII, which is almost 
imperceptible in component A.
To analyze this effect quantitatively, we calculated the maximum fluxes in
the CIV, \ciii, HeII, NIII lines after subtracting the continuum for
the components A and B (see Table \ref{fluxes}). Using these values, 
we also determined the flux ratios in the two components and the 
relative B/A ratio (see Table \ref{ratios}).

The ratio of amplitudes in the HeII emission line  
in each component is B/A$\sim 2$. In the case of NIII, B/A $\sim 35$.
These differences are clear in Figure 1, as is a very small
shift in all the emission lines between the two spectra due to the
redshift difference.

\subsection{ UV Resonance Lines }

The CIV $\lambda$ 1550 \AA\ line profile structure found in both lensed images
indicates BAL outflow which provides further evidence that UM 425 is a lens. In
Figure 2,  OVI $\lambda$ 1033 \AA\ obtained in an earlier epoch of UM425 A with
IUE (\cite{mom96}) is plotted on a common velocity scale with the CIV profiles
from both lensed images in the z$_{QSO}=1.471$ rest frame. The blue wings of
both OVI and CIV show BAL structure in high ionization resonance that extends
to velocities V$_{BAL}\sim -4000$ \kms. However, we also note that the low
ionization species MgII $\lambda$ 2800 \AA\ is present only in emission.
Absorption blueward of the emission profile is attributed to telluric water
vapor. The lack of BAL structure in low ionization species UM 425 indicates
UM425 is a high ionization BAL quasar. The distinct absorption line structure
in both images strengthens the case that UM 425 is a lens because the
probability of finding two BAL quasars within $\sim$6 arcsec with very similar
line profile structure is extremely small.   
      
\subsection{Absorption Structure}

In Figure 1, a prominent broad absorption feature that is unrelated to the BAL
lines can be seen in both UM425 A\&B with a centroid wavelength of $\lambda$
2175 \AA\ (rest frame). This corresponds to a redshift of z$_{abs} \approx$
1.48, essentially in the rest frame of the quasar. The measured equivalent
width is W$_{\lambda}$= 16.6 \AA\ with a full base that depresses the continuum
over $\sim$238 \AA. This feature corresponds to extinction that is commonly
seen in galactic sources (\cite{sm79}), but is not generally observed in QSO
spectra. This absorption feature has not commonly seen quasar spectra and
warrants further investigation. It may provide clues concerning the dust
environs of the BALQSOs, particulary if it reflects foreground absorption by a
dust torus or absorption in a host galaxy.  
 
The presence of the $\lambda$ 2175 \AA\ absorption feature provides a distinct
feature that enables us to determine the reddening of the QSO and allows us to
determine the continuous energy distribution of the QSO. This is important in
order to determine the exact level of hydrogen ionization present in the BAL
region because it relates to the excess metal abundances which are typically
associated with BAL QSOs which indicate a high nitrogen abundance relative to
hydrogen, carbon and oxygen and other metals.

\subsection{Discussion}

If UM425 is lensed, the difference in redshift between z$_A =1.471 \pm
0.002$ and z$_B = 1.477 \pm 0.002$ from the emission lines of images
A\&B must be reconciled with similar complex line profile structure
which is present in high-ionization ionic species. 
These observations established BAL line profile structure in image B,  that
follows from the line profile structure in high the resonance lines of CIV
$\lambda$ 1550. The BAL trough absorption is consistent with earlier IUE
observations in the far-UV of UM425A that revealed high velocity outflow in OVI
$\lambda$ 1033 and NV $\lambda$ 1240 \AA. The blue edges of the BAL troughs in
the far-UV lines in UM425 A indicate outflow with speeds that range up to $\sim
-12000$ \kms\ in OVI $\lambda$ 1033 and NV $\lambda$ 1240 \AA. The CIV BAL
trough presents a narrower velocity range that indicates outflow at speeds that
range up to $\sim-4000$ to $-5000$ \kms.

The measured differences in redshifts and amplitudes of several strong
emission lines between images A \& B may result from the intrinsic
variations in the QSO.  The wide angular separation between images of
$\Delta \theta \sim $6\farcs5 suggests a time delay of $\Delta t \sim
1.7$ years. This timescale is comparable to that for observed changes
in BAL line profile structure of $ \sim 1 $ year observed in UM425 A
with IUE (\cite{mom96}). The emission components of the BAL lines are
intrinsically broad, and changes in BAL absorption troughs that affect
the blue wings of the line profiles may alter the observed wavelength
centroid at different epochs of activity. This argument is purely
speculative at present and further spectral monitoring is required to
investigate this effect.

The lack of BAL structure in low ionization species such as MgII $\lambda$ 2800
\AA\ indicates UM425 is a  reddened, high-ionization BAL quasar. The
continuous energy distributions of low ionization MgII BALQSOs are usually
affected by reddening from dust in the quasar (Tunshek et al. 1994);
correcting for dust extinction in these sources is required in order to
reconcile their continuum distributions with those of the high ionization
BALQSO counterparts.

The presence of the $\lambda$ 2175 \AA\ absorption feature in both images in
the rest frame of the quasar provides important information concerning the
environs of the QSO. It also provides an additional feature that is common to
both images and further evidence that UM425 is a lens. Reddening may result
from foreground absorption by the host galaxy in which UM 425 is embedded, or
from a pre-existing dust torus that surrounds the system. The presence of dust
surrounding the BAL region is also important because it can affect the covering
factor by scattering photons from the QSO, thus effectively increasing the
global covering factor (cf. Turnshek \etal\ 1994).  

Baldwin (1977) found an indication of $\lambda$ 2175 \AA\ absorption in
QSO spectra of PHL938, Ton490, 3C286 and CTA102. Applying the Savage
\& Mathis (1979) galactic extinction law to these quasars, moderate
extinction levels of E(B-V)$\sim$0.08 for PHL938 and an E(B-V) $ \sim
0.04$ for Ton490 were obtained. However, the majority of QSOs surveyed
did not show the presence of the $\lambda$ 2175 \AA\ feature, while moderate
to low levels of dust absorption were suggested from the observed
continuum distribution in the survey.

The continuous energy distribution of the low ionization BALQSO PG 0043+039 was
studied by Turnshek \etal\ (1994), who found reddening by dust at levels of
E(B-V)=0.14 that are similar to the extinction level found here. Correcting
with the SMC extinction law,  Turnshek \etal\ (1994) obtained HI column
densities $\sim$4.4x10$^{21}$cm$^{-2}$. However, the means by which the
formation of dust grains proceeds and the manner by which grains survive the
ionizing environment of the BAL region remains obscure. Shielding by the BAL
clouds and the geometry of BAL outflow very likely are key components for
explaining the presence of dust in these systems (\cite{vo93}).

We applied the galactic Savage \& Mathis (1979) law to the UM425 spectra based
upon the presence of the $\lambda$ 2175 \AA\ dust feature. An E(B-V)$\sim$0.1
was obtained, or A$_v$$\sim$0.3, which is similar to the extinction level found
for the low ionization BALQSO PG0043+339 by Turnshek \etal\ (1994). Without
correcting for dust extinction in UM425, the continuous energy distribution
between $\lambda\lambda 2045-3124$ \AA\ in the z$_{QSO}=1.47$ rest frame
indicates a power law of $\lambda F_{\lambda}$= -0.67. Correction with an
E(B-V) = 0.1 corresponds to a steeper spectral of $\lambda F_{\lambda}$ =
-1.76, which may reflect emission from a disc. 
  
\section{Summary and Conclusions}

We obtained MMT spectra of both images of the gravitational lens candidate UM
425 to determine the relative redshifts of lines in the respective
spectra, and to ascertain whether UM425 is lensed. We further explored the BAL
line profile structure in both images previously observed only in the brighter
image A (m$_V$$\sim$16.2) with IUE (m$_V$$\sim$21.8 for image B).

The MMT data confirmed BAL structure in the CIV $\lambda$ 1550 \AA\ 
line in both images, consistent with the source being
lensed. Although a cross-correlation of the spectra of A and B restricted to
the region containing the principal emission lines indicates differences in
redshift of z$_B$-z$_A$ = 0.005$\pm$ 0.001, based upon the centroid
wavelengths of  NIII, \ciii, HeII, SiIV and MgII, the difference may be
explained by the complex broad emission profile of BAL lines and changes in the
BAL trough that can affect the emission profile at different epochs of
observation, given the long delay time for the system of $\Delta$t$\sim$1.7
years for images A \& B. Temporal variability and the time delays for 
the 2 images may also explain the differences in line flux ratios
for A \& B shown in Table \ref{ratios}. However, more detailed monitoring
is required to confirm this possibility. 

The MMT spectra of both images in UM425 indicate the presence of dust
extinction, as we deduce from the presence of the $\lambda$ 2175 \AA\
absorption feature observed in the rest frame of the quasar.  Such a 
strong absorption feature in both images is strong evidence in favor
of UM 425 being lensed, because the probability of finding 2 
high-ionization BAL quasars within $\sim$6 arcsec with very similar
line profile structure is negligible.

We would like to thank Craig Foltz for establishing the connection
among us.  AGM would like to thank E. Dwek for useful discussions.
JAM is supported by a postdoctoral fellowship from Ministerio de
Educaci\'on y Cultura, Spain.  We also thank the anonymous referee for
correcting an error in the first version of our manuscript.

\newpage
\begin{figure}[h]
{\epsfxsize=16cm \epsfbox{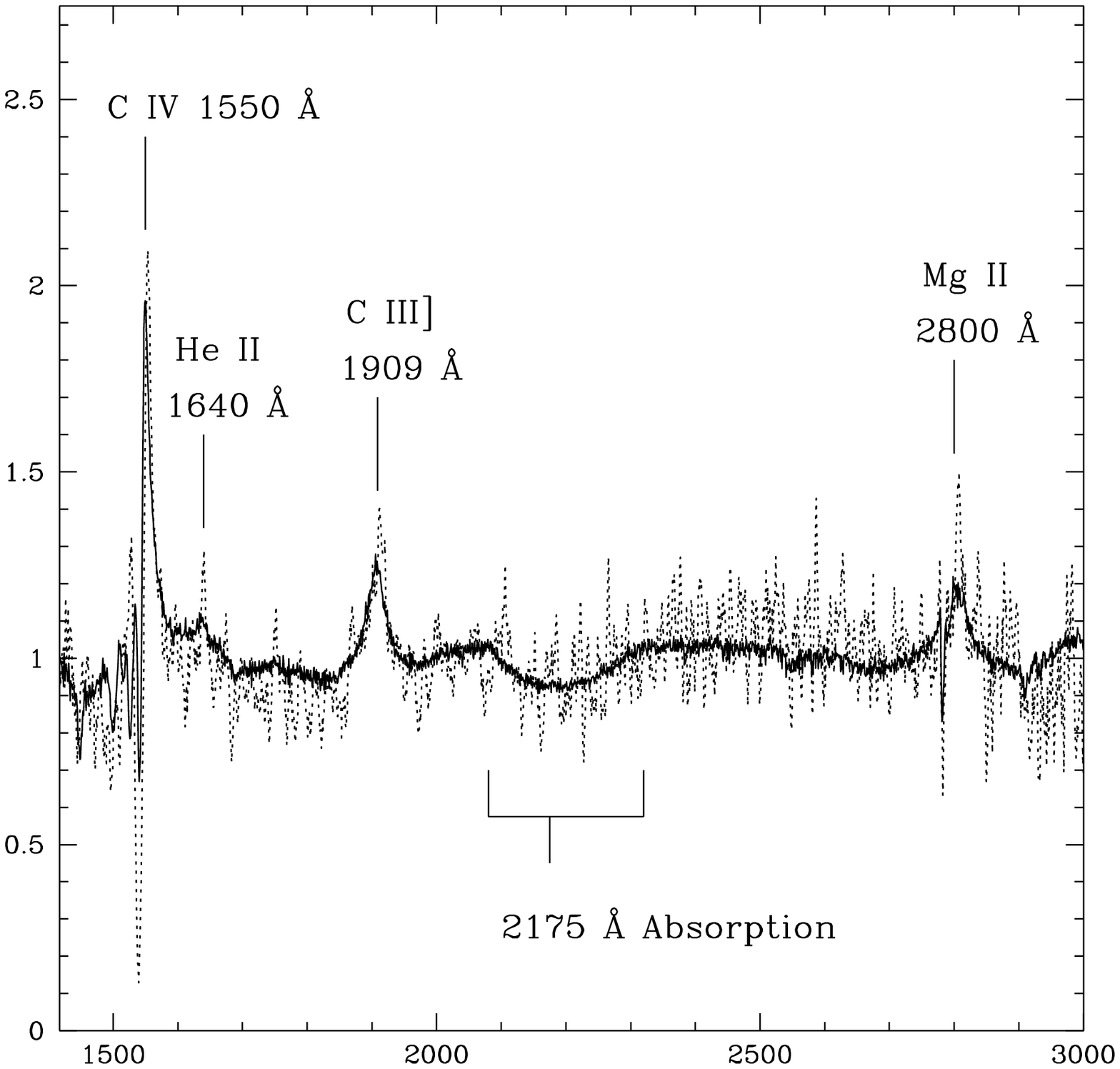}}
\figcaption{\small Calibrated spectra of the quasar UM425 
(components A \& B). 
Because of the large difference in luminosity between the two components, the
spectra of A and B are normalized to
a common scale for display purposes. The solid (dashed) line 
corresponds to the A (B) component. The abscissa shows observed wavelengths.
Prominent emission lines, as well as the $\lambda$ 2175 
\AA\ absorption are labeled.}
\label{fig1}
\end{figure}

\begin{figure}[h]
{\epsfxsize=16cm \epsfbox{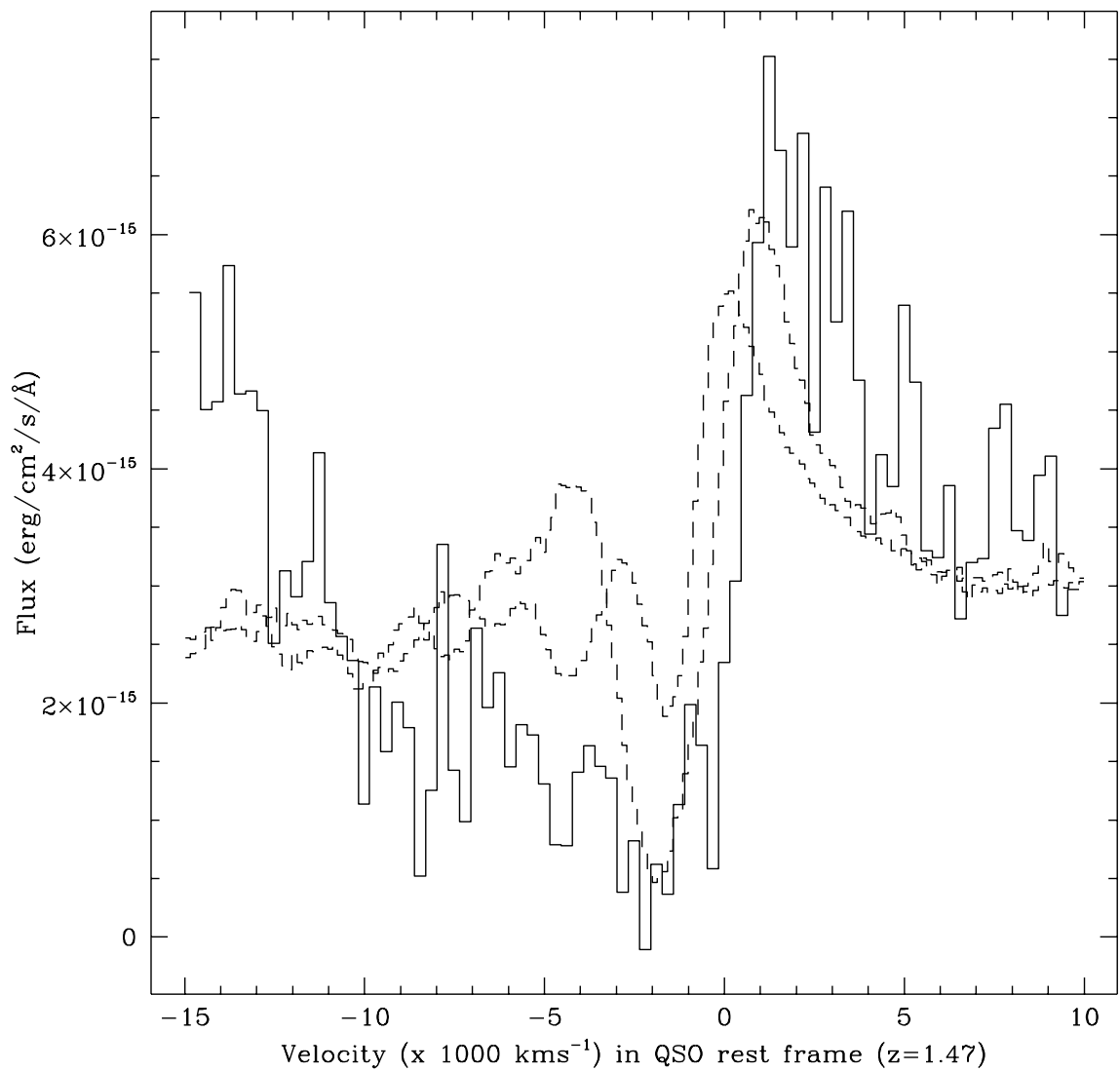}}
\figcaption{\small The OVI $\lambda$ 1033 \AA\ line profile (solid histogram) 
obtained from the IUE satellite (LWP29924) on 1995 Feb 7.6 
(JD 2449,756.1) of UM425 A
is plotted on a common velocity scale in the QSO rest frame,
together with the CIV $\lambda$ 1550 \AA\ obtained of UM425 A (dashed upper
curve) and UM425 B (lower dashed curve).  The fluxes for CIV
were scaled to a common level for display purposes.
The abscissa shows velocities with respect to rest for component A.}
\label{fig2}
\end{figure}



\newpage
\begin{deluxetable}{clcrrccc}
\tablewidth{15.9cm}
\tablecaption{ MMT observations log for UM425, UT 10 February 1997}
\tablehead{
No. & Object  &
Central $\lambda$& 
\multicolumn{1}{c}{Exp.} & 
\multicolumn{1}{c}{UT}&  
\multicolumn{1}{c}{airmass} & 
\multicolumn{1}{c}{PA} & 
\multicolumn{1}{c}{seeing FWHM}\nl
         &         & 
\multicolumn{1}{c}{\AA} &  
\multicolumn{1}{c}{sec}    &   &       &
\multicolumn{1}{c}{$^{\circ}$ E of N} &
\multicolumn{1}{c}{$''$}
}
\startdata
1   &    UM425    &   6000  & 2700&  7:54:04 &  1.25 & -28.9 & 2.12\nl
2   &    UM425    &   6000  & 2700&  9:40:51 &  1.16 & -28.9 & 2.15\nl
3   &    GB191B2B &   6000  &   30&  1:53:57 &  1.10 &   0.0 & 1.58\nl
4   &    PG1545+035&  6000  &   30& 13:14:25 &  1.15 &   0.0 & 2.12\nl
\tableline
\enddata
\label{tab1}
\end{deluxetable}
\begin{deluxetable}{lll}
\tablewidth{12.cm}
\tablecaption{ Line fluxes}
\tablehead{
 & \multicolumn{1}{c}{A}&\multicolumn{1}{c}{B}
}
\startdata
CIV  &  1.494 $10^{-15}$ & 2.705 $10^{-17}$  \nl
\ciii &  4.002 $10^{-16}$ & 7.878 $10^{-18}$  \nl
HeII &  1.880 $10^{-16}$ & 7.645 $10^{-18}$  \nl
NIII &  3.894 $10^{-18}$ & 2.633 $10^{-18}$  \nl
\tableline
\enddata
\tablecomments{\small Maximum fluxes after subtracting 
the continuum in components A and B, in erg/sec/cm$^2$/\AA.}
\label{fluxes}
\end{deluxetable}

\begin{deluxetable}{lllr}
\tablewidth{12.cm}
\tablecaption{ Line flux ratios}
\tablehead{
& \multicolumn{1}{c}{A}&\multicolumn{1}{c}{B}&  B/A
}
\startdata
\ciii/CIV  &  0.27   &  0.29  &  1.07  \nl
HeII/CIV  &  0.13   &  0.28  &  2.15  \nl
NIII/CIV  &  0.0026 &  0.097 & 37.30  \nl
HeII/\ciii &  0.47   &  0.97  &  2.06  \nl
NIII/\ciii &  0.0097 &  0.33  & 34.02  \nl
\tableline
\enddata
\tablecomments{Flux ratios for components A and B, and the ratio B/A.}
\label{ratios}
\end{deluxetable}

\end{document}